
%
%
\input epsf
\magnification=1200
\baselineskip=24pt plus 4pt minus 4pt
\null
\vskip 1.0in
\centerline{\bf ABSTRACT}
\bigskip
\bigskip
Most of the present understanding of the S=1 quantum spin chains displaying the
Haldane gap is coming from the so-called valence-bond-solid (VBS)
Hamiltonian which has an exactly known ground state. We show that this point
is characterized by the onset of short-range incommensurate spin correlations
in the one-parameter family of Hamiltonians
$ H_\theta = \cos\theta \sum_{i} {\bf S}_i \cdot {\bf S}_{i+1} +
\sin\theta \sum_i ({\bf S}_i \cdot{\bf S}_{i+1} )^2 $.
This gives a physical meaning to this special point.
We establish precise values for the gaps,
correlations, the string order parameter, and identify the VBS point as a
disorder point in the sense of classical statistical mechanics.
It is a quantum remnant of the classical transition between a ground state with
long-range N\'eel order and a ground state with incommensurate long-range
order.
\vfill
\eject
\noindent{\bf I. INTRODUCTION}
\medskip
It has been conjectured by Haldane$^1$ that antiferromagnetic quantum spin
chains
have a disordered ground state with a gap to spin excitation when the spins
are integer. This phenomenon has been studied extensively over the years
and a simple physical picture has emerged through the consideration of
the so-called valence-bond solid (VBS) Hamiltonian$^2$. This peculiar
Hamiltonian
contains, in addition to the simplest isotropic bilinear
nearest-neighbor exchange, a
biquadratic term in the case of the spin S=1 chain. This modification
transforms
the Hamiltonian in a sum of projection operators and, as a consequence, the
ground
state is known exactly and has a simple structure. This is different from the
exactly integrable models: here nothing is exactly known about the excited
states.
It is believed that this model is smoothly connected to the usual
nearest-neighbor
antiferromagnet: they share the same physics. More precisely, there is a hidden
topological long-range
order$^3$ that is common to both Hamiltonians$^{4,5}$ and which is revealed
clearly in the VBS Hamiltonian. It is interesting to note that a similar
situation
happens also in the fractional quantum Hall effect$^6$: here the Laughlin wave
function which is the exact ground state of an approximate Hamiltonian does
possess the hidden order which is revealed in the anyonic gauge. The VBS nature
of
the ground state of the S=1 spin chain has led to the curious consequence of
effective spins S=1/2 at the end of open chains: this has been observed
theoretically$^7$ by numerical means and experimentally$^{8,9}$.

If we concentrate on the VBS model in the $S=1$ case, it
can be written as the sum of a bilinear and a biquadratic spin-spin interaction
between nearest neighbors. It is thus natural to study it as a special case
of the general bilinear-biquadratic isotropic quantum $S=1$ chain:
$$
H_\theta = \cos\theta \sum_{i} {\bf S}_i \cdot {\bf S}_{i+1} +
\sin\theta \sum_i ({\bf S}_i \cdot{\bf S}_{i+1} )^2 ,
\eqno(I.1)
$$
with $\theta$ varying between 0 and $2\pi$. All energies are measured in units
of
the global exchange coupling which is omitted everywhere in this paper.
The VBS Hamiltonian corresponds to the value $\theta_{vbs}$ with $\tan
\theta_{vbs} =1/3$. In this case,  each term in the sum in Eq.(I.1) ${\bf S}_i
\cdot {\bf S}_{i+1} +({\bf S}_i \cdot{\bf S}_{i+1} )^2 /3$ is the projector on
the
spin S=2 state of the two neighboring spins $i, i+1$. This fact leads to a
simple
ground state wavefunction$^2$. The behaviour of this model as a function of
$\theta$ has been studied by numerous authors$^{10-30}$. If one increases
$\theta$ starting from the bilinear Hamiltonian $\theta =0$
which is known to possess a
Haldane gap there is no phase transition till $\theta =\pi/4$ and thus the VBS
Hamiltonian ($\theta_{vbs}={\rm atan} (1/3)<\pi/4$) is smoothly connected to
the
usual bilinear Heisenberg model. However its precise physical meaning has
remained so far unexplained.

In this paper, we clarify the physical meaning of the VBS point in the phase
diagram of the family (I.1) of models. Consider first the classical limit
$S\rightarrow\infty$ of Eq.(I.1). For $\theta =0$ the ground state is
antiferromagnetically long-range ordered with ordering wavevector $q=\pi$
(N\'eel
state). When $\theta$ is increased, the order becomes incommensurate when
$\theta >\theta_c$ with $\tan \theta_c =1/2$: the wavevector shifts from
$q=\pi$.
As a consequence, the static structure factor $S(q)$ has a delta peak at
$q=\pi$
when $\theta <\theta_c$ and a delta peak at $q<\pi$ when $\theta >\theta_c$.
Now,
in the quantum case $S<\infty$ with S {\it integer}, fluctuations wash out
long-range order and we are left with short-range order below a characteristic
correlation length $\xi$, according to Haldane's conjecture. The delta peak of
$S(q)$ is thus smeared and acquire a finite width given by
$\xi^{-1}$.

It is important to note that, due to this finite width, the
incommensurate behaviour cannot be seen immediately in the quantity $S(q)$
when $q$ shifts away from the commensurate position. This is best
understood by considering the analytic structure of $S(q)$ in the complex
q-plane.
Short-range order means that singularities (poles or branch points) are away
from
the real axis at a distance $\approx \xi^{-1}$. In the commensurate phase, the
real part of the nearest singularity is $\pi$ and the peak of $S(q)$ for q real
is also at $q=\pi$. If we increase the parameter $\theta$, at some value the
real
part of the leading singularity will move away from $q=\pi$: this means that
real-space correlations oscillate with a new period. However, due to the width
of
the peak in the structure factor, the maximum of $S(q)$ remains at $q=\pi$ till
the shift $\Delta q$ of the real part reaches a value $O(\xi^{-1})$. Then for
larger values of $\theta$, the structure factor will exhibit an incommensurate
peak. We have obtained evidence that, right at the VBS point, the correlations
become incommensurate in real space: it is a "disorder" point in the language
of
classical statistical mechanics$^{31}$. For a {\it larger} value of $\theta$,
the
function $S(q)$ exhibits the signature of incommensurability$^{29}$: this point
is properly called a Lifshitz point. This splitting of the classical
phenomenon  at $\theta_c$ is typical of systems with only short-range
order.

We proceed by first recalling briefly in section II the state of knowledge on
the
bilinear-biquadratic $S=1$ quantum spin chain. In section III we import the
concepts of so-called "disorder" points from classical statistical mechanics.
They are discussed by use of simple classical spin models.
In section IV, using the  Density Matrix Renormalization Group
algorithm$^{32}$,
we calculate energy gaps, correlation functions and correlation lengths, as
well
as the $S=1$ string order parameter in the neighborhood of the VBS point. We
demonstrate that the spin correlations exhibit a change of behaviour in real
space right at the VBS point and that this point is a quantum example of a
"disorder" point. Section V contains our conclusions.
\vfill
\eject

\noindent{\bf II. PHASE DIAGRAM OF THE BILINEAR-BIQUADRATIC SPIN CHAIN}
\medskip

Let us represent the phase diagram of models $H_\theta$ as in figure 1. For
$\theta=0$ and $\theta=\pi$, one finds the isotropic (anti)ferromagnetic
quantum
Heisenberg  model. Antiferromagnetic (resp. ferromagnetic) models corresponds
to
$-\pi/2<\theta<+\pi/2$ (resp. $+\pi/2<\theta<3\pi/2$).
Some points in the phase diagram have been studied in detail and we summarize
below the current knowledge:
\bigskip
\noindent
$\bullet$ $\theta=0$: Isotropic antiferromagnetic quantum Heisenberg model:
this well-studied model has a non-degenerate disordered ground state with
exponentially decaying antiferromagnetic  correlations obeying a law $\langle
{\bf S}_0 \cdot {\bf S}_n\rangle\approx (-)^n\exp(-n/\xi)/\sqrt{n}$ and a
gapped
spectrum (Haldane gap $\approx 0.41$). The static structure factor $S(q)$ is a
square-root Lorentzian peaked at $q=\pi$.
\bigskip
\noindent
$\bullet$ $\theta=0.1024\pi$ ($\tan\theta = 1/3$): this is the VBS model with
exact valence-bond-solid ground state. The spin correlations are {\it purely}
exponential with a correlation length $\xi = (\ln 3)^{-1} \approx 0.91$. There
is
a gap in the spectrum ($\Delta = 0.664$).
\bigskip
\noindent
$\bullet$ $\theta=0.25\pi$: this is the Lai-Sutherland model, the Hamiltonian
is a
sum of permutation operators and exactly integrable by the Bethe
ansatz$^{10,11}$.  The ground state is unique and  the model is critical. The
corresponding conformal theory is
$SU(3)_{k=1}$. There are zero-energy modes for $q=0, \pm 2\pi/3$.
\bigskip
\noindent
$\bullet$ $\theta=-0.25\pi$: The model is solvable exactly by the nested Bethe
ansatz$^{12,13}$. One finds a critical system with a unique ground
state. The conformal theory is $SU(2)_{k=2}$. There are zero-energy
modes at $q=0,\pi$.
\bigskip
\noindent
$\bullet$ $\theta=-0.50\pi$: the physics is that of a dimerized state; the
order
parameter is given by the coefficient $c_2$ in the singlet-singlet correlation:
$$
\langle ({\bf S}_i {\bf S}_{i+1}) ({\bf S}_j {\bf S}_{j+1})
\rangle \rightarrow c_1 +(-1)^{i-j}c_2 ,
\eqno(II.1)
$$
for $|i-j|\rightarrow\infty$. The ground state is twice degenerate in the
thermodynamic limit and the spectrum is gapped ($\Delta=0.17$). The correlation
length is given as $\xi=42.2$: these are exact results$^{18-21}$.
\bigskip
\noindent
$\bullet$ $\theta=-0.75\pi$: possible location of a continuous phase transition
from a ferromagnetic to a dimerized phase$^{23,25,26}$.
\bigskip

\noindent
$\bullet$ $\theta=\pi$: This is the isotropic ferromagnetic Heisenberg model.
There is ferromagnetic order with  gapless excitations. The ground state
is the ferromagnetic state for $\pi/2<\theta <5\pi /4$.
\bigskip

With these points, one constructs the following phase diagram$^{14,17}$:
Starting at $\theta=\pi$, one finds an ordered ferromagnetic state without gap.
The ferromagnetic phase terminates at
$\theta=-0.75\pi$. A continuous phase transition leads to a dimerized
state. A prediction by Chubukov$^{22}$ of a non-dimerized
nematic phase seems refuted by Fath and Solyom$^{25,26}$. In the dimerized
phase, the ground state is a singlet with a double
degeneracy due to a $Z_2$ symmetry breaking. The order parameter is given by
$c_2$
in the correlation function (II.1). A continuous phase  transition at
$\theta=-0.25\pi$ leads to a Haldane phase, with a unique  disordered ground
state, exponentially decaying correlations and a gapped spectrum. This gapped
phase ends at the Lai-Sutherland point $\theta =+0.25\pi$ where a continuous
transition leads to a phase which is possibly trimerized (see refs. 25,26 for a
detailed discussion). One is back to ferromagnetic phase for $\theta =+0.5\pi$.
This phase diagram is displayed in figure 1. Up to now, the VBS point appears
to
be generic in the Haldane phase.

Recently,  Bursill, Xiang and Gehring$^{29}$ considered, using the DMRG, the
Fourier transform of the spin-spin correlations, i.e. the static structure
factor:
$$
S(q)=\sum_n {\rm e}^{iqn}\langle {\bf S}_n \cdot {\bf S}_0  \rangle ,
\eqno(II.2)
$$
In the Haldane phase, at the isotropic point $\theta =0$, $S(q)$ is a
square-root
Lorentzian with a peak at $q=\pi$. Since parity is unbroken in the phases we
discuss, we restrict the momenta to the interval $(0,\pi)$.
It was found that the peak of the Fourier transform $S(q)$ starts to move away
from $q=\pi$ towards $q=\pm 2\pi/3$. This happens at $\tan \tilde{\theta} =
0.43806(4)$, or $\tilde{\theta}=0.1314\pi$. They found also that when
$\theta\rightarrow 0.25\pi$, the peak reaches $2\pi /3$ in agreement with
the period-3 zero modes that are seen  at the Lai-Sutherland point.
Their conclusion is then that there are three regions between $\theta=0$ and
$\theta=0.50\pi$:
\bigskip
$\bullet$ $0 < \theta < \tilde{\theta}$: short-ranged antiferromagnetic
correlations.
\bigskip
$\bullet$ $\tilde{\theta} < \theta < 0.25\pi$): short-ranged {\it spiral}
order;
the peak of the Fourier transform shifts from $q=\pi$ to $q=2\pi/3$; the
spectrum
is still gapped.
\bigskip
$\bullet$ $0.25\pi\leq \theta < 0.50\pi$: possible trimerized phase beyond the
Lai-Sutherland phase transition.
\bigskip

If one considers the classical limit of model (I.1), there is a related
phenomenon. For $\theta$ smaller than $\theta_c=\arctan (1/2)=0.148\pi$,
the ground state  is the usual commensurate N\'eel order with wavevector
$q=\pi$,
while beyond this value of $\theta$ the ground state becomes an incommensurate
spiral characterized by wavevector $q$ such that $\cos q=-{1\over 2} {\rm
cotan}
\theta$. Of course the classical ground states have long-range ordering and
when
going to finite spin values this order becomes short range. We will show in
section IV that this short-range order naturally splits the
commensurate-incommensurate transition in {\it two} distinct phenomena: one
happens at $\theta_{vbs}$ where the spin oscillations becomes incommensurate in
real space and one happens at
$\tilde\theta$ where incommensurability becomes obvious in the structure
factor.

Before discussing our results, we  now recall the corresponding concepts of
classical statistical physics, first developed by Stephenson$^{31}$.

\bigskip
\noindent{\bf III. SHORT-RANGE ORDER AND "DISORDER" POINTS}
\medskip

If one starts from a classical model and considers finite integer spins then,
according to Haldane's conjecture, there is only short-range order and a finite
correlation length. This is, roughly speaking, an example of "quantum
paramagnetism". Finite spin is in a sense equivalent to a finite temperature.
In Haldane's mapping$^1$ onto a nonlinear sigma model, the coupling constant is
equal to the inverse of the spin while in the nonlinear sigma model describing
classical two-dimensional systems the coupling is the temperature itself.
This means that an integer spin chain has a physics which is related to that of
a
two-dimensional spin system at nonzero temperature. Since there is no
long-range ordering
in
such a two-dimensional system according to the Mermin-Wagner theorem, the
ground
state of the spin chain is short-range ordered. We consider thus classical
systems
in their paramagnetic phase to understand the physics of finite-spin chains.
Strictly speaking, one should consider classical two-dimensional systems but
in fact, for our purposes, the physics is absolutely similar to that of
three-dimensional systems above the critical temperature.

Let us consider a magnetic Hamiltonian that exhibits two ordered
low-temperature
phases, one with commensurate correlations and the other with
incommensurate correlations. One may think for example of a square lattice of
classical spins with nearest-neighbor exchange $J_1$ and third-nearest-neighbor
$J_3$: when $J_3/J_1>1/8$ one destabilizes the N\'eel order and obtains an
incommensurate spiral whose pitch evolves continuously. We note
$P$ any parameter that controls the zero-temperature phase transition
(e.g. anisotropy, pressure, ratio of exchange couplings, etc.). A generic phase
diagram is given in figure 2. We consider the
case where these low-temperature phases are separated from the disordered
paramagnetic high-temperature  phase  by {\it continuous} transitions (If the
classical system is 2D then the $T_c$ is zero and the reasoning is unchanged).
It
is  clear  that the short-ranged correlations in the disordered phase
will be of variable nature:  ``close'' to the commensurate phase, they will be
commensurate; ``close'' to the  incommensurate phase,
they will be incommensurate. One can guess that there will be a
line in this phase diagram, where the correlations change their behaviour; this
change  will be linked to correlations of very short range, thus to a state
with a
minimum of  short range order. Hence the name of disorder line.
If one moves along path (A) in the paramagnetic phase in figure 2, there should
be a change in the correlations.

If one considers the real-space spin-spin correlations along path (A) they will
develop incommensurate oscillations at some point $A_D$.
If one considers now the correlations in Fourier space $S(q)$, one finds that
the
peaks of the Fourier transform  still stays at the value for commensurate
correlations, even though the real-space  correlations are already
incommensurate,
due to the finite correlation length:  the peak width is linked to $\xi^{-1}$.
It
is only ``closer'' to the incommensurate  phase that the peak will
start to shift. This will happen at a second point $A_L$ on path (A) in fig.2.
This is
easy to understand by taking a simplified form for $S(q)$:
$$
S(q)= {1\over \alpha (q-q_x)^2 +(q-q_x)^4 +\xi^{-2}}.
\eqno(III.1)
$$
For convenience, we shift the momenta to set $q_x =0$. It is only
for $\alpha <0$ that $S(q)$ has a double-peak structure. When $\alpha >0$ is
large enough, $\alpha^2 >4\xi^{-2}$, all the poles of (III.1) are on the
imaginary axis in complex-q space and the real-space correlations do not
oscillate. But when
$\alpha^2 <4\xi^{-2}$ the poles have a real part and thus there are real-space
oscillations. It is only when the real part of the poles is large enough that
the structure factor itself displays a two-peak shape. This effect is entirely
due to the finite correlation length $\xi$ i.e. short-range ordering and finite
width of the peak in $S(q)$. It is clear from the simple example above that it
is
only the analytic structure of $S(q)$ that matters and not our peculiar
eq.(III.1). As such, this is a general behaviour.

The starting point of real-space oscillations is the disorder point. It extends
in the plane of Fig.2 in a disorder line D. The starting point for the double
peak
structure in $S(q)$ is the Lifshitz point, extending in a line L in fig.2. In
experiments, one normally measures the structure factor in reciprocal space and
will thus observe this line. It is also clear that the two lines must end in
the multicritical point where the three phases meet, which is thus necessary
for
their existence.

To avoid unnecessary generalizations, we take the results from an example
treated by an RPA method in Ref.33. It is an Ising spin chain with a
ferromagnetic
$J_1$ interaction between nearest neighbors and an  antiferromagnetic $J_2$
between next-to-nearest neighbors. The RPA treatment shows that there are
three
regimes.  The disorder temperature is:
$$
T_D = {J_1^2}/{4|J_2|} + 2|J_2|.
\eqno(III.2)
$$
One derives the following expressions for $x$ large:
\bigskip
$T < T_D$: $\langle S(0)S(x) \rangle \simeq e^{-x/\xi_-(T)}$ .
\bigskip
$T = T_D$: $\langle S(0)S(x) \rangle \simeq x {\rm e}^{-k_0x}$ with $\cosh k_0a
=
J_1/4|J_2|$.
\bigskip
$T > T_D$: $\langle S(0)S(x) \rangle \simeq {\rm e}^{-x/\xi_+(T)}
\cos (q(T)x)$
with $q(T) \sim (T-T_D)^{1/2}$.
\bigskip
For $\xi_{\pm}(T)$ one finds that, on the commensurate side, the correlation
length  exhibits an infinite derivative at $T_D$; it will
typically be very small, but not necessarily a minimum or zero. The derivative
on the incommensurate side is finite (see figure 3). This characterizes a
disorder
line of the first kind.  There are two more special properties: (i) The
susceptibility shows  a particularly simple form at the
disorder line. (ii) If one considers the correlation functions and compares
them
to an  Ornstein-Zernicke correlation function (for $x$ large) for a
$d$-dimensional system:
$$
\langle S(0) S(x) \rangle \simeq e^{-x/\xi(T)}/r^{(d-1)/2} ,
\eqno(III.3)
$$
one sees that the correlation functions are those for $d=1$, as expected
for a chain, except at the disorder point: formally, they correspond to $d=-1$.
Let us add that the incommensurate correlations are given by a wave vector $q$
which  shifts continuously from the commensurate value $q=0$; the exponent
$1/2$
is however non-universal.

In the spin chain problem with S=1, we are always in the paramagnetic phase
i.e.
we are following a path like (A) in fig.2 when varying the parameter $\theta$
in
the model (I.1). We thus expect to cross the disorder point and the Lifshitz
point
that are the quantum remnants of the classical transition at $\theta_c$.

\bigskip
\noindent{\bf IV. THE NEIGHBORHOOD OF THE VBS POINT}
\medskip
For our calculations, we use the DMRG:  see Ref.32 for a detailed discussion of
the algorithm.  We apply it to chains of a length $L=96$ and keep $M=80$
states.
This is sufficient to find truncation errors smaller than $10^{-12}$ in the
considered region. It is therefore not necessary to extrapolate results in $M$,
as
they are extremely close to the exact results.
For the VBS point, we recover the exact results within machine precision.
Due to the very small correlation lengths (typically smaller
than 3) a length $L=96$ is sufficient to obtain the results of the
thermodynamic limit. From the DMRG viewpoint, this situation is ideal. There is
however a problem with purely computational errors: the spin-spin correlations
are, for a distance of 30 to 40 sites, of the order $10^{-13}$ or less.
As they are obtained by summing  small numbers below machine precision
(in a REAL*8 calculation), they must be rejected. To judge the importance of
this
effect, we have adopted the following strategy: as the system under study is
isotropic and disordered in the region where the Haldane and the trimerized
phase
meet, the spin-spin correlations must obey the relation
$$
\langle S^+_iS^-_j \rangle =
\langle S^x_iS^x_j + S^y_iS^y_j \rangle = 2 \langle S^z_i S^z_j \rangle.
\eqno(IV.1)
$$
These two quantities are calculated independently; we reject correlations that
show  a deviation of more than a thousandth from this relation.
As a matter of fact, we find
that the correlations $\langle S^z_i S^z_j \rangle$ reach a minimum value which
oscillates randomly around $10^{-14}$, whereas $\langle S^+_i S^-_j \rangle$
continues to diminish regularly. We conclude that the values for
the latter correlation are more
precise; analyzing the calculation, we find that for the latter all the
important
contributions and weights show the same sign, whereas it changes for the
former.

The key of our analysis is  not to analyze the Fourier transform of the
correlations, but to analyze them directly in real space.
We will thus show that the VBS point, so far without special role
in the phase diagram, is effectively a disorder point. It shows all the
characteristics of a disorder point of the first kind, as described in section
III.

Consider the real-space correlations (figure 4) for some values of $\theta$
between $0.10\pi$  and $0.125\pi$. This includes the VBS point
($\theta_{vbs}=0.1024\pi$).  The correlations for $\theta<0.1024\pi$ are
perfectly
antiferromagnetic; this is most evident in a logarithmic plot of
$|(-1)^{n}\sqrt{n} \langle {\bf S}_n \cdot {\bf S}_0\rangle|$.
For antiferromagnetic correlations, the curve shows no modulations. Above
$\theta_{vbs} = 0.1024\pi$, the logarithmic plots show oscillations with
periods
that become shorter for increasing $\theta$, to end at a period of 3 for
$\theta\rightarrow 0.25\pi$. Thus the VBS point is a disorder point.
The correlations are no longer antiferromagnetic, but
already  incommensurate: this point is missed if one considers only the Fourier
transform as in Ref.29.
These modulations can be understood from the classical law for an
incommensurate high-temperature phase (adapted to $d=2$):
$$
\langle  {\bf S}_n \cdot {\bf S}_0\rangle \approx
\cos [q(\theta)n] {{{\rm e}^{-n/\xi(\theta)}}\over {\sqrt{n}} }= (-1)^n
\cos [(\pi-q)n]{ {{\rm e}^{-n/\xi(\theta)}}\over {\sqrt{n}}}.
\eqno(IV.2)
$$
The modulations should thus show a period $\pi/(\pi-q)$, which is easier to see
than the period originating directly from $\cos qn$. To show that the
correlation
functions can be well described by (IV.2), we have attempted a direct fit of
our results. This fit is complicated by the fact that there are
effectively three parameters to be controlled, the wave vector $q$,
the correlation length $\xi$ and also a phase factor $\phi$, from replacing
$n=i-j$ by $(n-\phi)$ in the argument of the cosine. We find that the fit is
extremely sensitive to the parameter values, which allows for a good fit.
As an example, we take $\theta=0.115\pi$, sensibly below the point given so
far for the change of the correlations. We obtain the fit shown in figure 5,
for
$\phi=0.65$, $\xi=1.08$ and $\pi-q=0.198\pi$. The wave vector $q$ has already
shifted by 20 percent from the antiferromagnetic value $q=\pi$.

For $\theta$ closer to the VBS point, the fit is made more complicated by the
errors in the correlation function due to the finite precision of the computer.
To estimate the behaviour $q(\theta)$, we therefore consider the periodicity:
For a period $p$, $\pi - q \approx \pi/p$. The discrete nature of the problem
limits this approach. We find  the behaviour shown in figure 6. Clearly, is
is  compatible with $q \propto (\theta-\theta_{vbs})^{\sigma}$
with $0<\sigma<1$, the behaviour of a disorder point of the first kind.
The curve would be compatible with $\sigma \approx 1/2$; but we are in no
position to give a precise estimate.

We have also calculated the Fourier transform of the correlation function and
find
results in agreement with those given by Bursill et al.$^{29}$.
The VBS point has no special significance for its behaviour; the point
$\tilde{\theta}=0.1314\pi$  where the peak starts to shift, can now be
identified
as a Lifshitz point: see fig.7. As expected in a system with short-range order,
it
is distinct from the disorder (VBS) point.

For the correlation lengths, we find that they show a minimum for the VBS
point,
with an infinite slope (numerically: very large) for $\xi$ in
the commensurate regime ($\theta < 0.1024\pi$), and a slow increase in the
incommensurate regime with a finite slope, given in figure 8. The correlation
lengths have been found by different methods: in the regime $\theta <
\theta_{vbs}$,  we have compared the spin-spin correlations numerically and
graphically to a law $\exp (-n/\xi) /\sqrt{n}$, which were in all cases in good
agreement. We estimate  the precision of the results of the order of 1 per
cent:
the truncation errors are of the  order $10^{-13}$, a serious underestimation
can therefore be  excluded. For $L=96$ and $\xi \approx 1 - 2$, finite size
effects are of no importance.  The situation is more
complicated for the regime $\theta > \theta_{vbs}$.
As we have seen, the fit of the
theoretically expected behaviour to the found curve is rather complex. If one
considers a plot of
$|(-1)^{n}\sqrt{n} \langle {\bf S}_i \cdot {\bf S}_j \rangle|$ (figure 4),
one finds that a linear fit for the maxima is quite good. This is stable in
the sense that a factor $\cos qx$ influences the logarithm
least when it is close to 1. Very generously estimated, the error of the
graphical evaluation should be below 5 per cent. We estimate that for $\theta >
0.15\pi$ the underestimation due to a non-negligible truncation error
dominates.
For $\theta_{vbs} < \theta \leq 0.11\pi$, we could not obtain the correlation
length: on  the one hand, the periods due to the incommensurability are too
long to separate them well  from the exponential behaviour. On the other hand,
neither a fit $\exp -r/\xi$ nor $\exp (-r/\xi)/\sqrt{r}$
is satisfactory. We don't know whether this observation is
due to the crossover of the
behaviour of the correlation function or simply due to problems of the
numerical
method. In any case, the results indicate strongly an extrapolation to
the VBS point with a finite slope.

The other important quantities are the gap and the string order
parameter $O_{\pi}(i,j)=\langle S^z_i \exp (i\pi \sum_{k=i+1}^{j-1} S^z_k)
S^z_j\rangle$.
The gap shows a maximum for $\theta \approx 0.123\pi$, which is
thus linked neither to the disorder nor the Lifshitz point: see figure 9.
For the  transition points $\theta=\pm 0.25\pi$ we have not obtained serious
estimates: the critical  fluctuations imply a greater $M$, and the vanishing
gap is difficult to see. Our results are well compatible with a zero gap, but
not
precise enough to give a serious estimate.  The
point $\theta=-0.20\pi$ is sufficiently close to the transition to cause the
same problem. For the VBS point, we
find a gap value $\Delta=0.664$, in agreement with Ref.25. Since the gap is
smooth at $\theta_{vbs}$, this implies that the spin wave velocity has a
singularity at this point ($c=\Delta\xi$).

The string order parameter in the thermodynamic limit
$i-j\rightarrow\infty$ has its  extremum for the VBS point: $|O_{\pi}| =
4/9$: see figure 10. One sees that the VBS
point, though a point of minimal {\it spin} order, is a point of maximum
hidden
topological order.

To complete the identification of the VBS point as a disorder point, we note
that there is the equivalent of the particularly
simple form of the susceptibility  of the classical model of section III. The
exact correlations at the VBS point obey a one dimensional Ornstein-Zernicke
form: they are purely exponential (no prefactor),
whereas the non-linear sigma model yields two-dimensional  correlation
laws. This ``dimensional reduction'' is accompanied by a  particularly simple
form
of the Hamiltonian: it can be decomposed  into a sum of local projection
operators. The problem loses  its quantum character and turns into a classical
one-dimensional problem.

\bigskip
\noindent{\bf V. CONCLUSION}
\medskip
We have shown that the VBS point$^2$ is a disorder point in the sense of
classical
statistical mechanics. It is thus identified as a point which is not just by
chance exactly solvable, but shows this property for more profound physical
reasons. The results of Bursill et al. who have considered the Fourier
transform
of the spin-spin correlations fit naturally in the picture we have developed:
the
point they identify as the point  where correlations
change is simply the Lifshitz point following the definition given above.
Quantum fluctuations in the integer spin chain wash out long-range order. As a
consequence, the transition that happens at the classical level for $\theta_c$
is no longer a phase transition. However the change of short-range correlations
still happens in the quantum system at the VBS point in real-space. Due to the
finite correlation length this is not seen immediately in $S(q)$, hence the
Lifshitz point which is distinct. Since $\xi\rightarrow\infty$ when the spin
increases one expects that these two points should merge in the classical limit
right at $\theta_c$. This is summarized in figure 11.

This implies the following description of the phase diagram: There are  three
regions for $0 < \theta < 0.25\pi$:

$0 < \theta < \theta_{vbs}$:
Short-range antiferromagnetic correlations;
Haldane phase with gapped spectrum. A generic description is given by the
VBS model or also the isotropic antiferromagnetic Heisenberg model.

$\theta_{vbs} < \theta < \tilde{\theta}$:
Incommensurate short-range
correlations with a wave vector $q<\pi$, that shifts away from $\pi$ as
$\pi - q \propto (\theta-\theta_{vbs})^{\sigma}$, $\sigma \approx 1/2$.
In the Fourier transform,
the peak stays at $q = \pi$. The spectrum is gapped: we expect that the
low-lying
Haldane modes are now at the incommensurate wavevector. Part of the VBS physics
remains valid:  the gap, the hidden order, and the  free spins 1/2 at
the ends of an open chain.

$\tilde{\theta} < \theta < 0.25\pi$: The physics is similar to that of
the former region; the peak of the Fourier transform shifts from $q = \pi$
to
$q=\pm 2\pi/3$ and the incommensurate correlations become visible. The spectrum
is
gapped. There is however no profound physical difference between this region
and the former one.

The above picture does not challenge conventional wisdom in the sense that the
usual Heisenberg Hamiltonian $\theta =0$ shares the same physics with the VBS
Hamiltonian and there are no additional phase transitions between the
point$^{12,13}$ $\theta =-0.25\pi$ and the Lai-Sutherland point $\theta
=+0.25\pi$. However the VBS point itself means that the physics has changed
beyond
$\theta_{vbs}$: due to the incommensurate correlations, it is no longer
possible
to capture the low-lying excitations by a nonlinear sigma model with symmetry
breaking pattern
$O(3)/O(2)$ as used originally by Haldane. In the incommensurate regime, the
full
rotation group is broken down. Appropriate nonlinear sigma models have been
contemplated before$^{34}$. Since they involve a nonabelian symmetry, they are
generically massive, consistent with the gapped nature of the Haldane phase.

There is an interesting relationship$^{6}$ with the fractional Quantum Hall
effect. The present understanding$^{35}$ of the physics of the FQHE at filling
$\nu =1/m$, m odd, is based on Laughlin's wavefunction $\psi_m$ which is
the exact ground state of a truncated Hamiltonian. In addition, this function
embodies the hidden long-range order that is revealed in the anyonic gauge.
This is similar to the situation of the VBS wavefunction. Expectation values
computed with the Laughlin wavefunction correspond to a classical statistical
problem which is the two-dimensional one-component plasma (2D OCP): this is
Laughlin's plasma analogy. The case of the full Landau level $\psi_1$
corresponds to the special point$^{36}$ $\Gamma =2$ of the 2D OCP
($\Gamma$ being the ratio of the squared electric charge to the temperature)
at which the
density correlations begin to oscillate in real space, a precursor phenomenon
of the crystallization that occurs at $\Gamma \approx 140$ when the plasma is
dilute enough. This special point has also some of properties expected from a
disorder point$^{36}$. This is similar to what we observe at the VBS point.

\bigskip

{\bf Acknowledgements}
\bigskip
It is a pleasure to thank N. Elstner and O. Golinelli for fruitful discussions.

\vfill
\eject
\centerline{\bf FIGURE CAPTIONS}
\bigskip

\noindent
{\bf Figure 1}:

\noindent
Phase diagram of the bilinear-biquadratic $S=1$ isotropic
quantum spin chain as a function of $\theta$.
 Solid lines: transition points; dashed lines: other special points.
H(AFM):
isotropic antiferromagnet; VBS: the valence-bond-solid model; L: the
crossover point studied in Ref.29; H (FM): isotropic ferromagnet.
\bigskip
\noindent
{\bf Figure 2}:

\noindent
Schematic phase diagram: a disordered high-$T$ phase is linked by
two continuous transitions to two ordered low-$T$ phases.
$P$ is a parameter which controls the nature of the ground state.
The dashed lines represent the disorder (D) and Lifshitz (L) lines, where the
behaviour of the  correlations changes in real and in Fourier space
respectively.
\bigskip
\noindent
{\bf Figure 3}:

\noindent
Correlation length at a disorder point of the first kind (schematical
from RPA):
the commensurate phase is to the left.
\bigskip
\noindent
{\bf Figure 4}:

\noindent
Real-space spin-spin correlations as a function of the distance, for several
values of
$\theta$ below  and above the VBS point (at $\theta_{vbs} =0.1024\pi$).
The modulations appear above the VBS
point. Here $K_n ={\rm ln} \left[|(-)^n \sqrt{n}\langle {\bf S}_0\cdot
{\bf S}_n\rangle |\right]$.

\vfill
\eject

\noindent
{\bf Figure 5}:

\noindent
Comparison between the spin-spin correlations
predicted (solid line) and calculated numerically (squares + dashed line) for
$\theta=0.115\pi$,  above $\theta_{vbs}$, but below $\tilde{\theta}$.
The dotted line is $(-)^n \sqrt{n} \exp (n/\xi) \langle S^z_0 S^z_n\rangle$
and the solid line is $\cos \left[(n-\phi)(\pi-q)\right]$.
\bigskip
\noindent
{\bf Figure 6}:

\noindent
Wave vector $\pi - q$ characteristic of the spin correlations. There is a
singularity at the VBS point consistent with the identification of a disorder
point.
\bigskip
\noindent
{\bf Figure 7}:

\noindent
The Fourier transform $S(q)$ as a function of momentum for various values of
$\theta$. The two-peak structure appears only for $\theta>\tilde\theta
=0.1314\pi$ while
nothing
is seen at $\theta_{vbs}$.
\bigskip
\noindent
{\bf Figure 8}:

\noindent
Correlation lengths for various values $\theta$. The minimum is at the VBS
point.
\bigskip
\noindent
{\bf Figure 9}:

\noindent
Gaps in the Haldane phase as a function of $\theta$. The maximum value is
between
the VBS point and the Lifshitz point.
\bigskip
\noindent
{\bf Figure 10}:

\noindent
String order parameter $|O_{\pi}|$ in the thermodynamic
limit in the Haldane phase. The maximum is exactly at the VBS point.
\bigskip
\noindent
{\bf Figure 11}:

\noindent
Schematic phase diagram of integer spin chains with biquadratic coupling:
model (I.1). Here
the inverse spin plays the role of a temperature: the S=1 case corresponds
to short-range ordered phases. Of course this picture may be altered by other
types of ordering such as dimerization.

\vfill
\eject
\centerline{\bf REFERENCES}
\bigskip
\item{[1]}F. D. M. Haldane, Phys. Rev. Lett. {\bf 50}, 1153 (1983); Phys.
Lett. A{\bf 93}, 464 (1983).
\medskip
\item{[2]}I. Affleck, T. Kennedy, E.H. Lieb and H. Tasaki, Phys. Rev. Lett.{\bf
59}, 799 (1987); Comm. Math. Phys. {\bf 115}, 477 (1988).
\medskip
\item{[3]}M. den Nijs and K. Rommelse, Phys. Rev. {\bf B 40}, 4709 (1989).
\medskip
\item{[4]}T. Kennedy, J. Phys. C: Cond. Matter, C{\bf 2}, 5737 (1990).
\medskip
\item{[5]}T. Kennedy and H. Tasaki, Phys. Rev. B{\bf 45}, 304 (1992).
\medskip
\item{[6]}S. M. Girvin and D. P. Arovas, Phys. Scr. T{\bf 27}, 156 (1989).
\medskip
\item{[7]}S. Yamamoto and S. Miyashita, Phys. Rev. B{\bf 48}, 9528 (1993).
\medskip
\item{[8]}M. Hagiwara, K. Katsumata, I. Affleck, B. I. Halperin and J. P.
Renard,
Phys. Rev. Lett. {\bf 65}, 3181 (1990); S. H. Glarum, S. Geshwind, K. M. Lee,
M.
L. Kaplan and J. Michel, {\it ibid.} {\bf 67}, 1614 (1991).
\medskip
\item{[9]}O. Avenel, J. Xu, J. S. Xia, M. F. Xu, B. Andraka, T. Lang, P. L.
Moyand, W. Ni, P. J. C. Signore, C. M. van Woerkns, E. D. Adams, G. G. Ihas, M.
W. Meisels, S. E. Nagler, N. S. Sullivan, Y. Takano, D. R. Talham, T. Goto and
N.
Fujiwara, Phys. Rev. B{\bf 46}, 8655 (1992).
\medskip
\item{[10]}G. V. Uimin, JETP Lett. {\bf 12}, 225 (1970).
\medskip
\item{[11]}C. K. Lai, J. Math. Phys. {\bf 15}, 1675 (1974); B. Sutherland,
Phys.
Rev. B{\bf 12}, 3795 (1975).
\medskip
\item{[12]}L. A. Takhtajan, Phys. Lett. {\bf 87A}, 479 (1982).
\medskip
\item{[13]}H. M. Babudjian, Phys. Lett. {\bf 90A}, 479 (1982); Nucl. Phys.
B{\bf
215}, 317 (1983).
\medskip
\item{[14]}I. Affleck, Nucl. Phys. B{\bf 265}, 409 (1986).
\medskip
\item{[15]}J. Oitmaa, J. B. Parkinson and J. C. Bonner, J. Phys. C{\bf 19},
L595
(1986).
\medskip
\item{[16]}J. S\'olyom, Phys. Rev. B{\bf 36}, 8642 (1987).
\medskip
\item{[17]}N. Papanicolaou, Nucl. Phys. B{\bf 305} [FS23], 367 (1988).
\medskip
\item{[18]}J. B. Parkinson, J. Phys. C{\bf 20}, L1029 (1987); {\it ibid.} C{\bf
21}, 3793 (1988)
\medskip
\item{[19]}M. N. Barber and M. T. Batchelor, Phys. Rev. B{\bf 40}, 4621 (1989).
\medskip
\item{[20]}A. Kl\"umper, Europhys. Lett. {\bf 9}, 815 (1989); J. Phys. A{\bf
23},
809 (1990).
\medskip
\item{[21]}E. S. S\o rensen and A. P. Young, Phys. Rev. B{\bf 42}, 754 (1990).
\medskip
\item{[22]}A. V. Chubukov,  J. Phys. Condens. Matter {\bf 2}, 1593 (1990);
Phys.
Rev. B{\bf 43}, 3337 (1991).
\medskip
\item{[23]}G. F\'ath and J. S\'olyom, Phys. Rev. B{\bf 44}, 11836 (1991).
\medskip
\item{[24]}R. F. Bishop, J. B. Parkinson and Y. Xian, J. Phys. Condens. Matter
{\bf 5}, 9169 (1993).
\medskip
\item{[25]}G. F\'ath and J. S\'olyom, Phys. Rev. B{\bf 47}, 872 (1993).
\medskip
\item{[26]}G. F\'ath and J. S\'olyom, J. Phys. Condens. Matter {\bf 5}, 8983
(1993).
\medskip
\item{[27]}Y. Xian, J. Phys. C{\bf 5}, 7489 (1993).
\medskip
\item{[28]}T. Xiang and G. Gehring, Phys. Rev. B{\bf 48}, 303 (1993).
\medskip
\item{[29]}R. J. Bursill, T. Xiang and G. Gehring, J. Phys. A{\bf 28}, 2109
(1995).
\medskip
\item{[30]}G. F\'ath and J. S\'olyom, Phys. Rev. B{\bf 51}, 3620 (1995).
\medskip
\item{[31]} J. Stephenson, Can.\ J. Phys.\ {\bf 47}, 2621
(1969); ibid. {\bf 48}, 1724 (1970); ibid.\ {\bf 48}, 2118 (1970);
 J. Math.\ Phys.\ {\bf 12}, 420 (1970).
\medskip
\item{[32]} S. R. White, Phys. Rev. Lett. {\bf 69}, 2863 (1992); Phys. Rev.
B{\bf 48}, 10345 (1993); S. R. White and D. A. Huse, Phys. Rev. B{\bf 48}, 3844
(1993).
\medskip
\item{[33]} T. Garel and J. M. Maillard, J. Phys. C: Solid State
Phys. {\bf 19}, L505 (1986).
\medskip
\item{[34]} T. Dombre and N. Read, Phys. Rev. B{\bf 39}, 6797 (1989).
\medskip
\item{[35]} See the articles by R. B. Laughlin and S. M. Girvin in ``The
Quantum Hall Effect'', R. E. Prange and S. M. Girvin editors, Springer-verlag,
New York, 1990.
\medskip
\item{[36]} B. Jancovici, Phys. Rev. Lett. {\bf 46}, 386 (1981).
\medskip
\vfill
\eject
\nopagenumbers

\hfuzz=5pt
\baselineskip 12pt plus 2pt minus 2pt
\centerline{\bf PHYSICAL MEANING OF THE AFFLECK-KENNEDY-LIEB-TASAKI}
\centerline{\bf S=1 QUANTUM SPIN CHAIN}
\vskip 24pt
\centerline{U. Schollw\"ock, Th. Jolic\oe ur and T. Garel}
\vskip 12pt
\centerline{\it Service de Physique Th\'eorique}
\centerline{\it C.E.  Saclay}
\centerline{\it F-91191 Gif-sur-Yvette, France}
\vskip 12pt
\vskip 1.0in
\centerline{Submitted to: {\it Physical Review B}}
\vskip 4.0in
\noindent May 1995

\noindent PACS No: 75.40.Mg, 75.10.Jm. \hfill SPhT/95-054

\vfill
\eject
\vfill
\epsfxsize= 15.0truecm
\centerline{\epsfbox{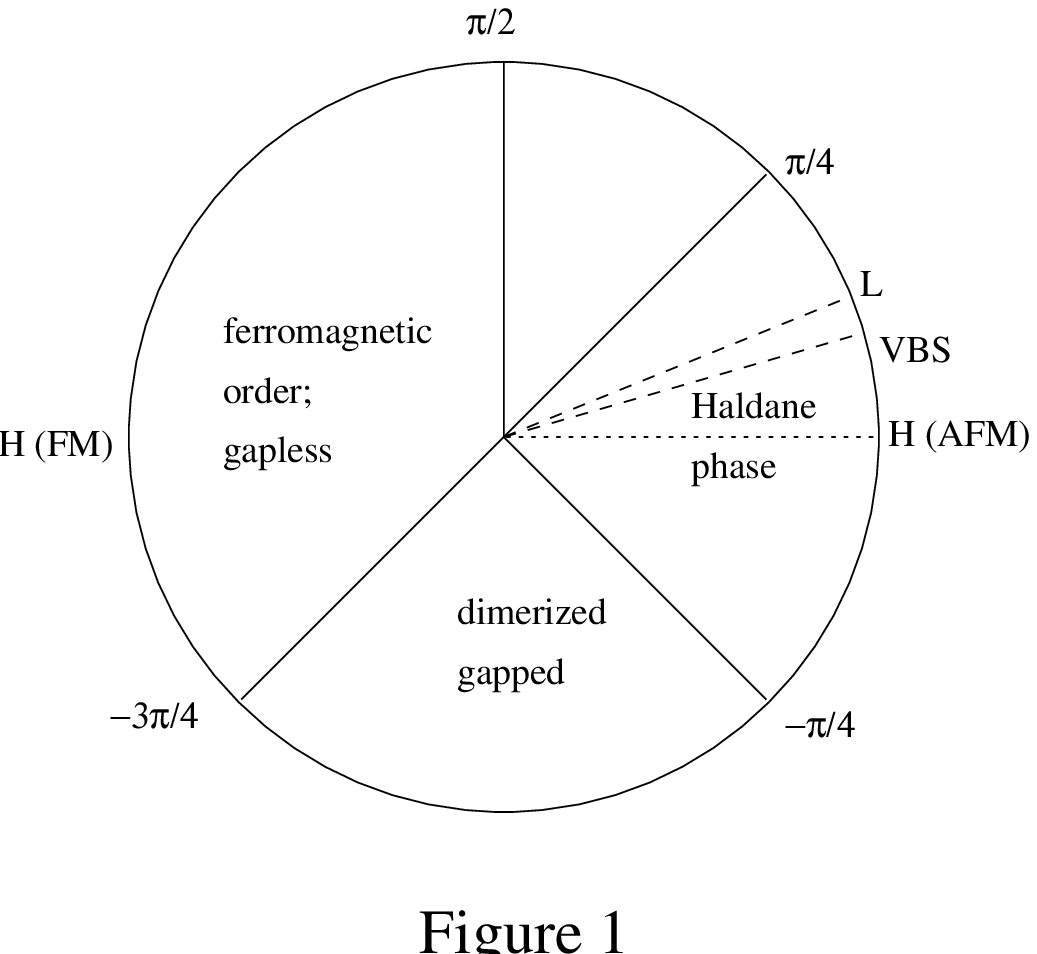}}
\vfill
\eject
\vfill
\epsfxsize= 15.0truecm
\centerline{\epsfbox{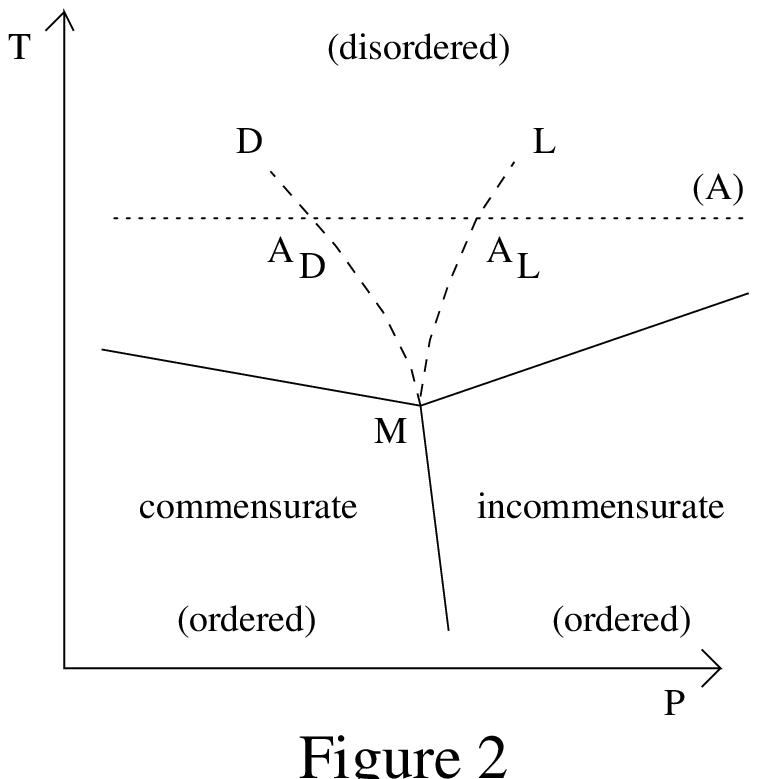}}
\vfill
\eject
\vfill
\epsfxsize= 15.0truecm
\centerline{\epsfbox{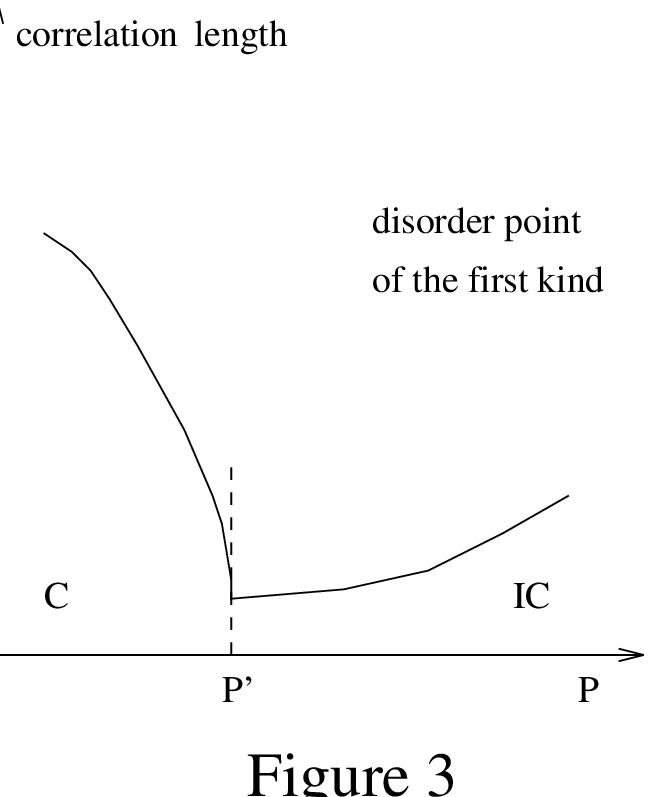}}
\vfill
\eject
\vfill
\epsfxsize= 15.0truecm
\centerline{\epsfbox{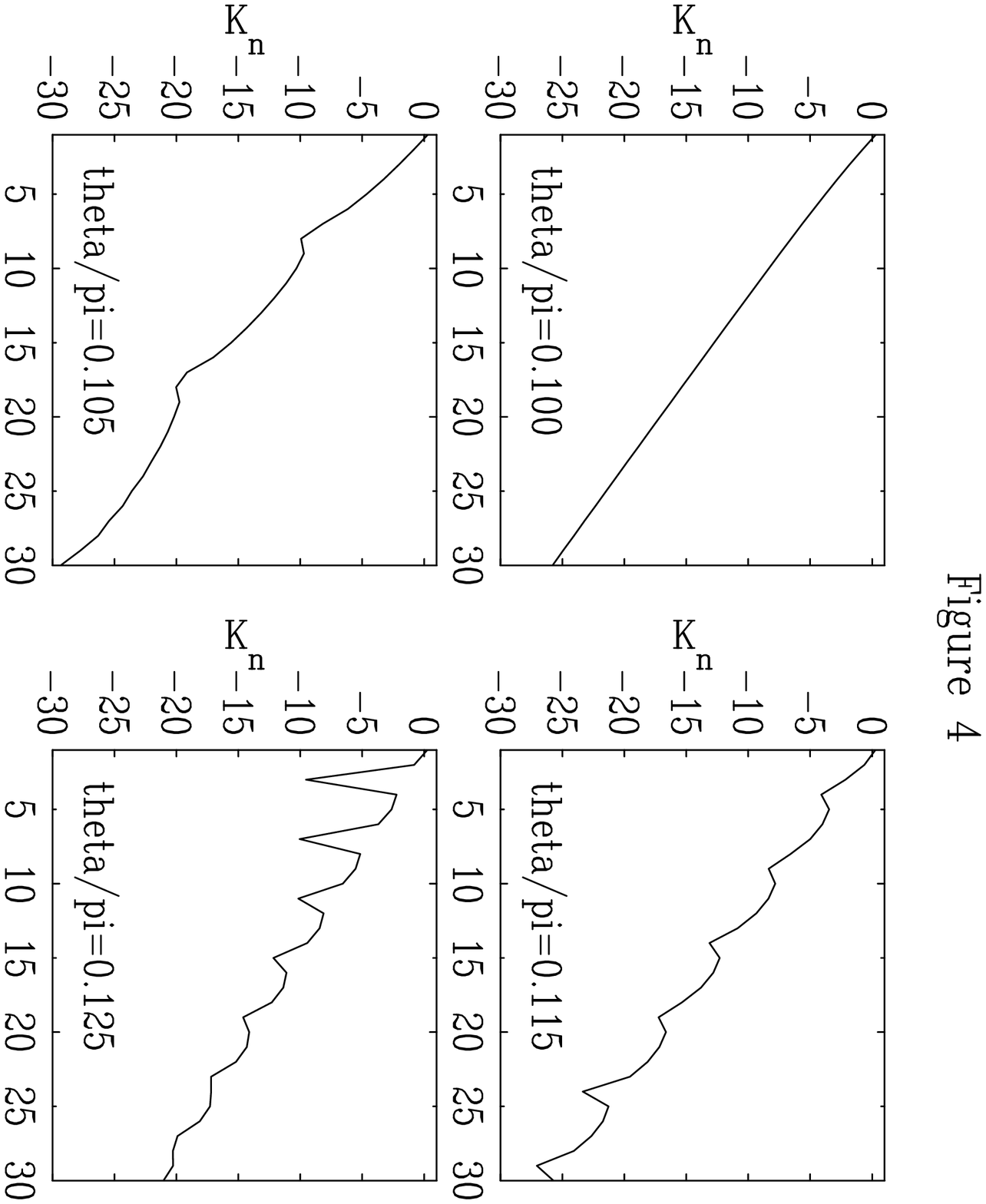}}
\vfill
\eject
\vfill
\epsfxsize= 15.0truecm
\centerline{\epsfbox{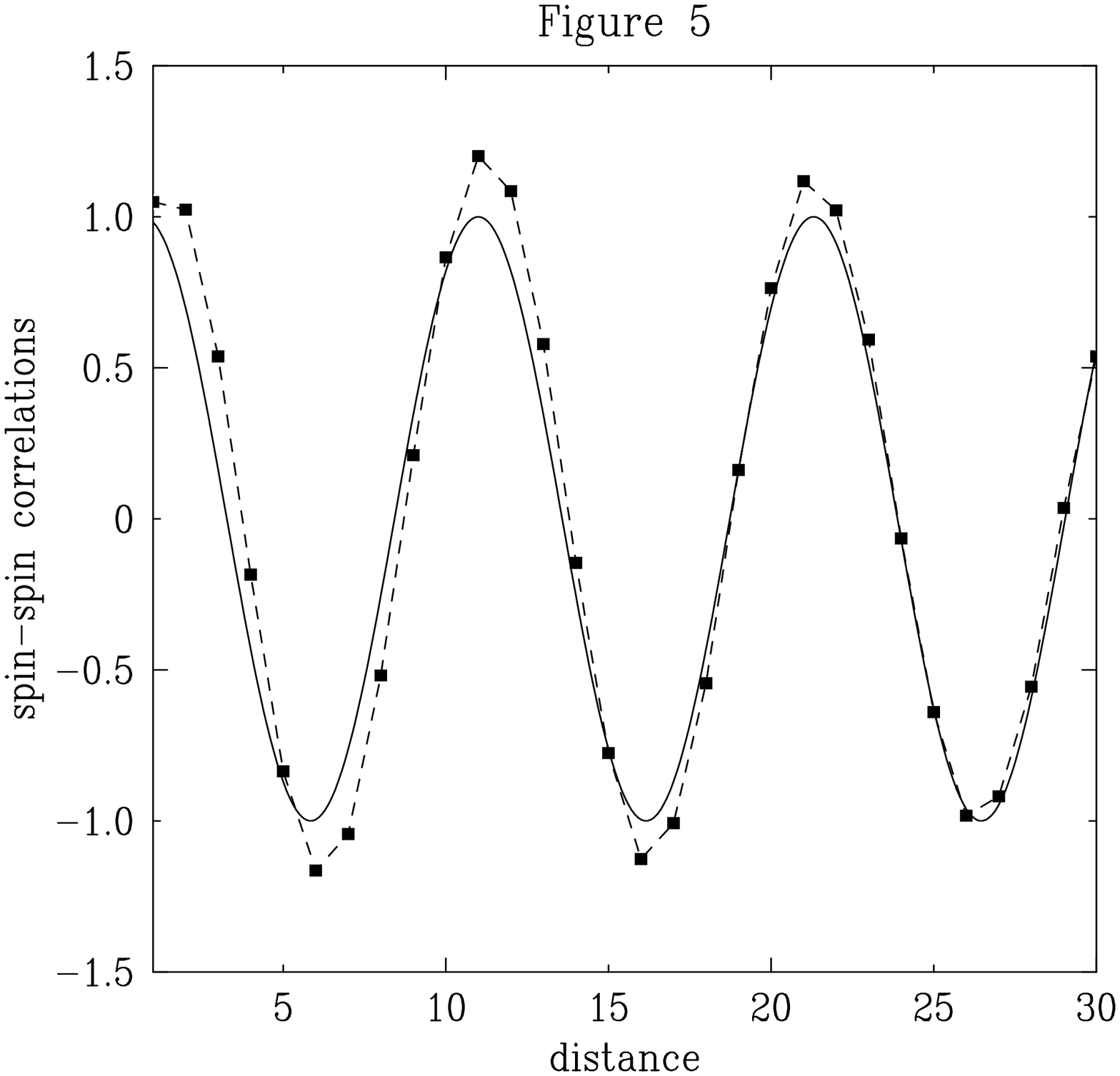}}
\vfill
\eject
\vfill
\epsfxsize= 15.0truecm
\centerline{\epsfbox{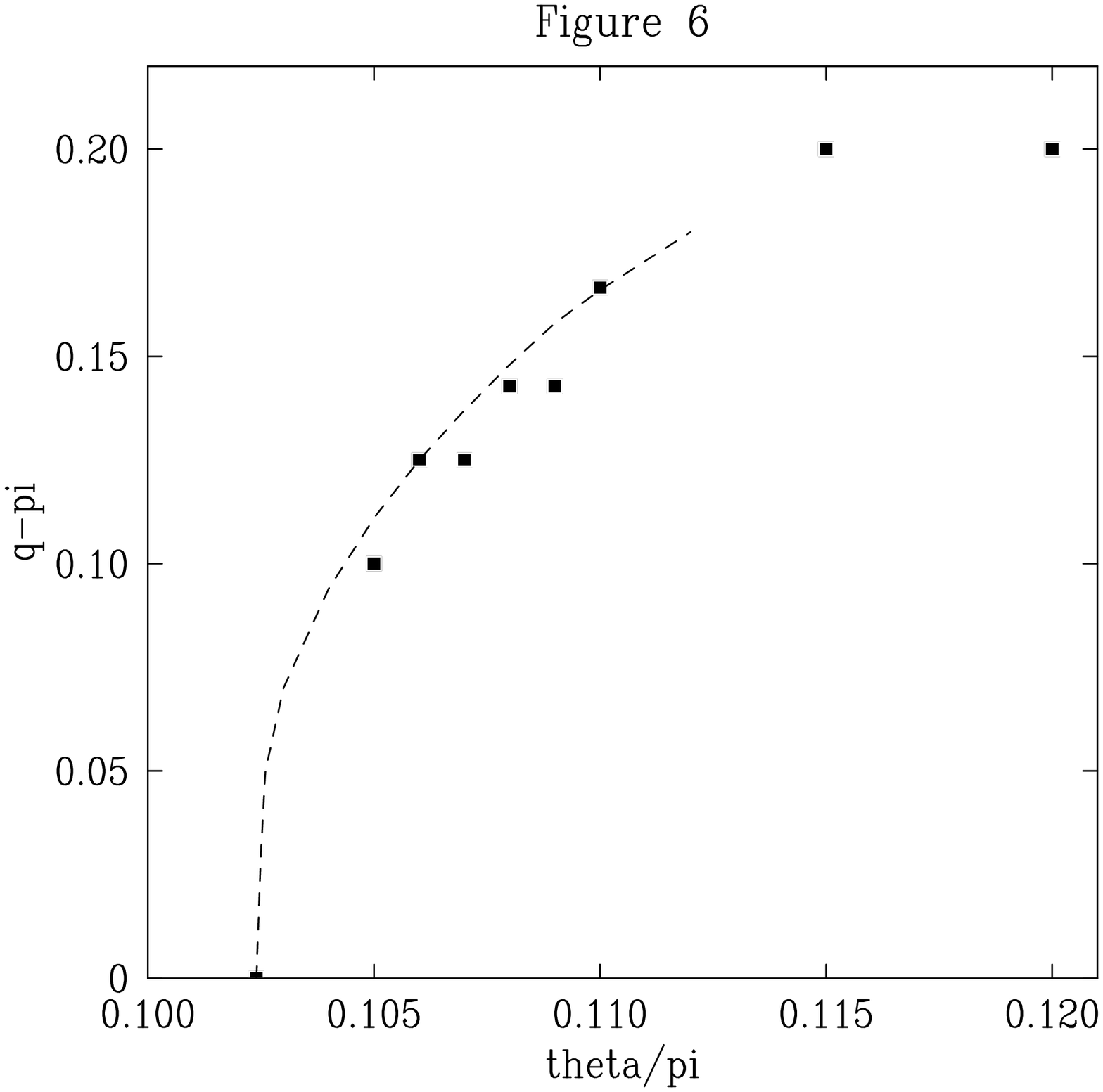}}
\vfill
\eject
\vfill
\epsfxsize= 15.0truecm
\centerline{\epsfbox{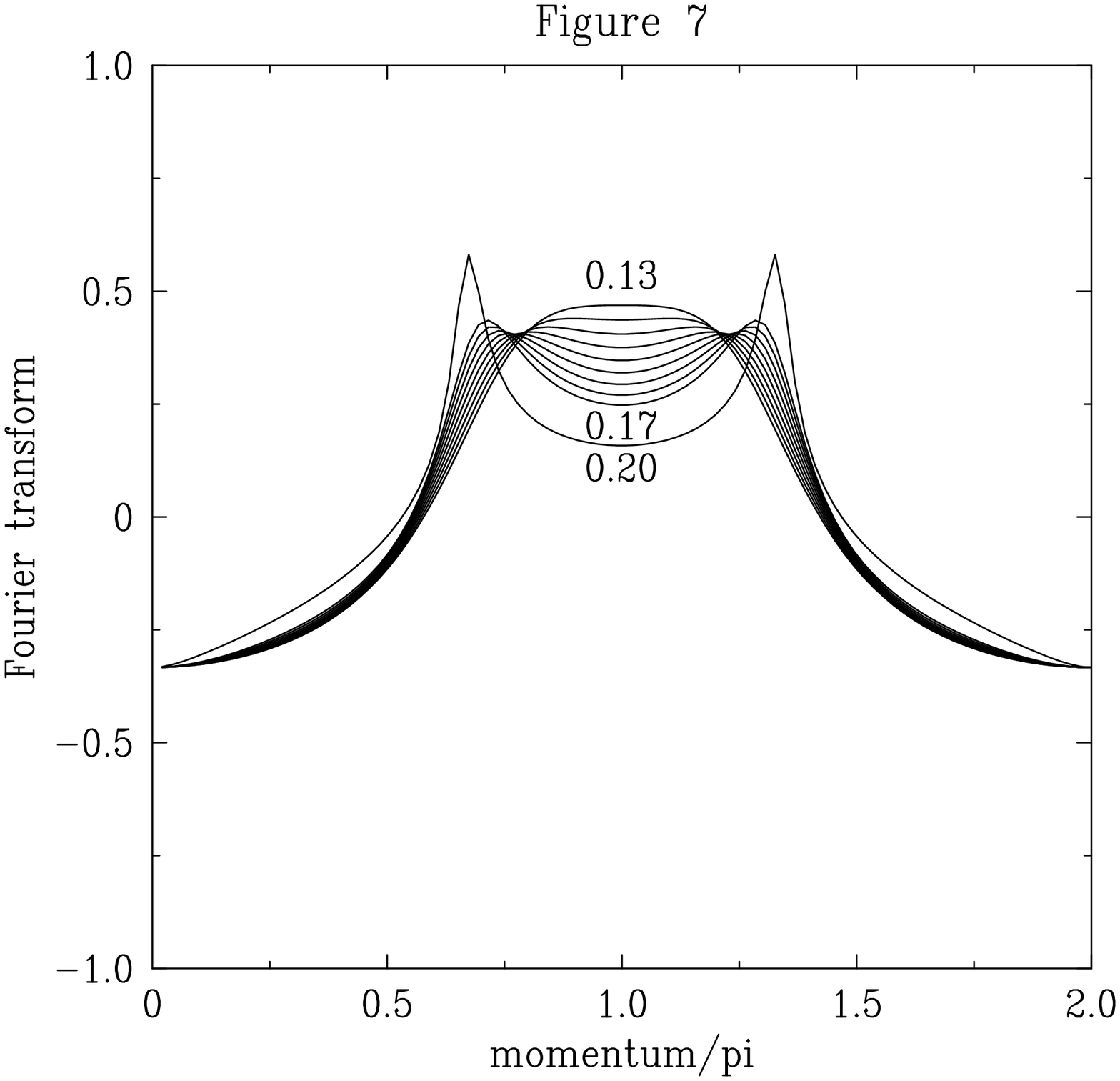}}
\vfill
\eject
\vfill
\epsfxsize= 15.0truecm
\centerline{\epsfbox{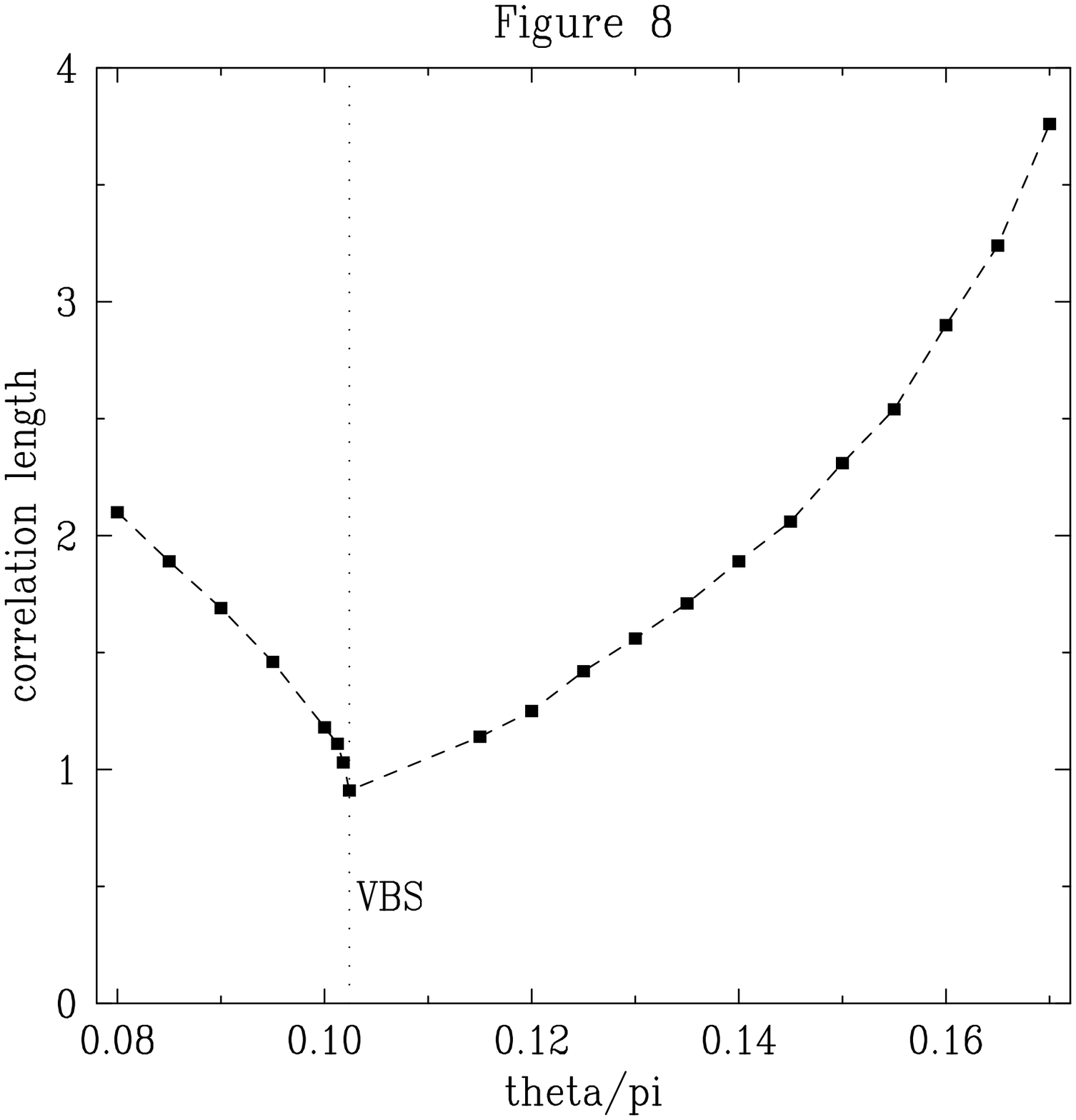}}
\vfill
\eject
\vfill
\epsfxsize= 15.0truecm
\centerline{\epsfbox{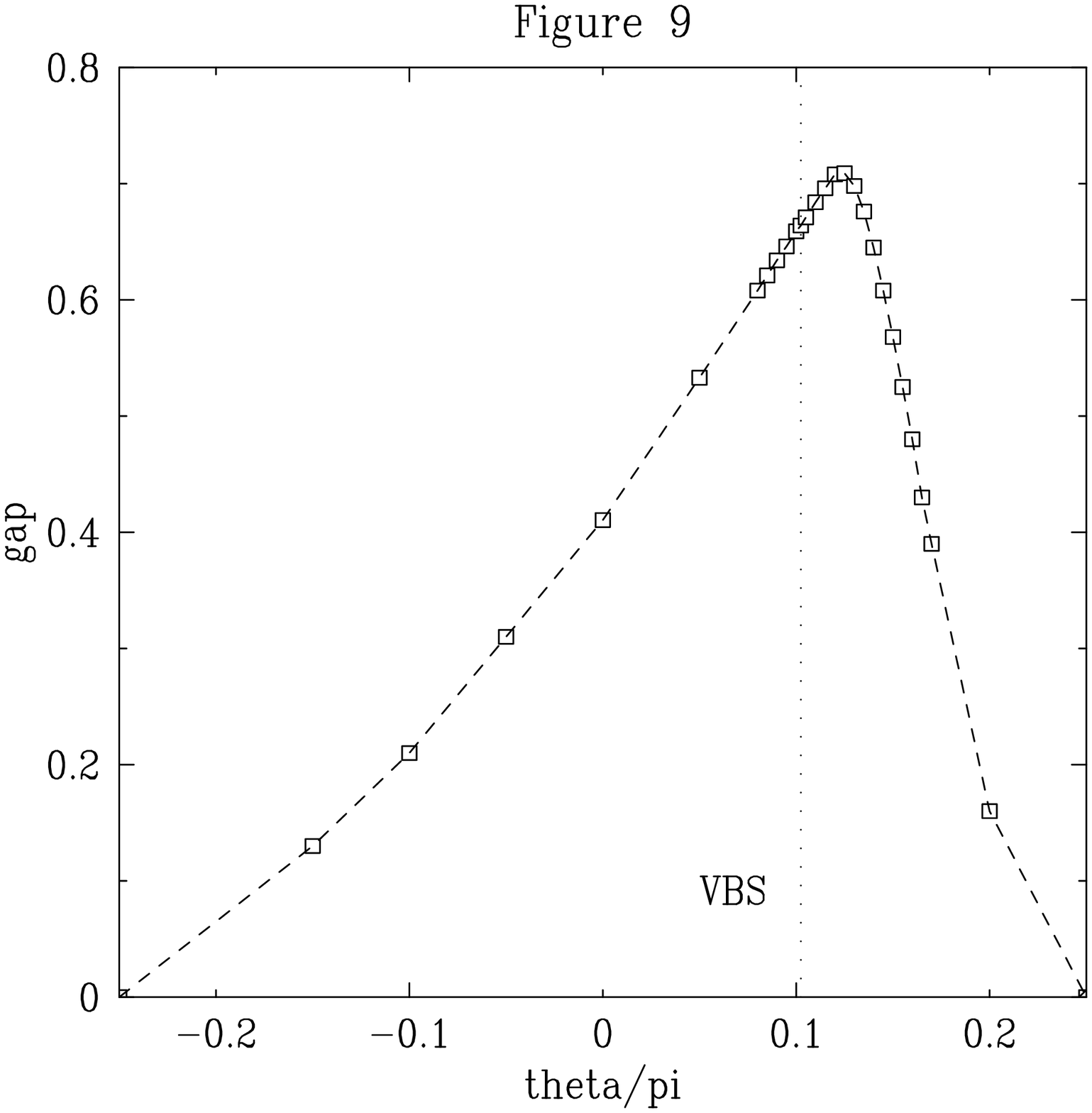}}
\vfill
\eject
\vfill
\epsfxsize= 15.0truecm
\centerline{\epsfbox{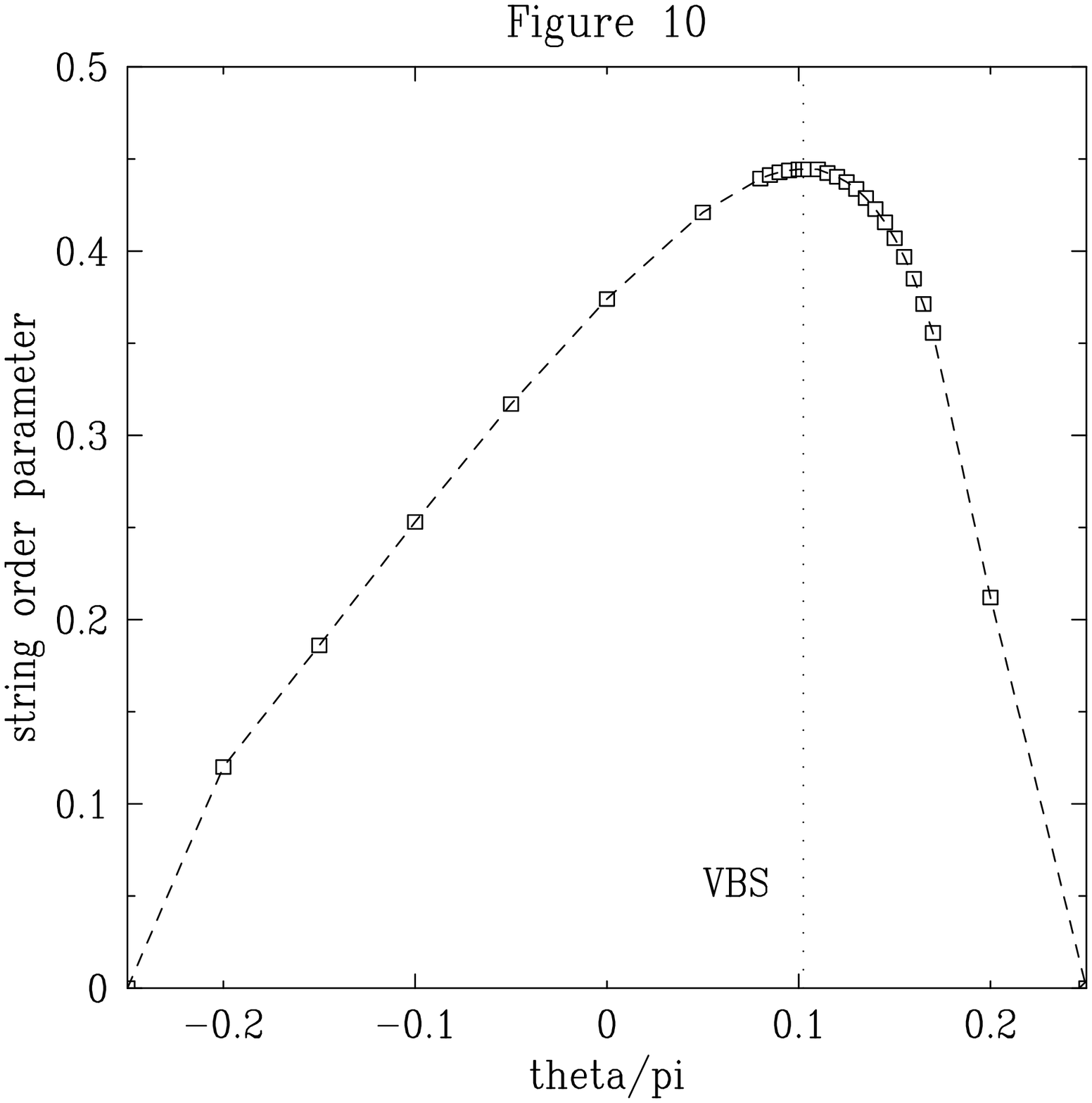}}
\vfill
\eject
\vfill
\epsfxsize= 15.0truecm
\centerline{\epsfbox{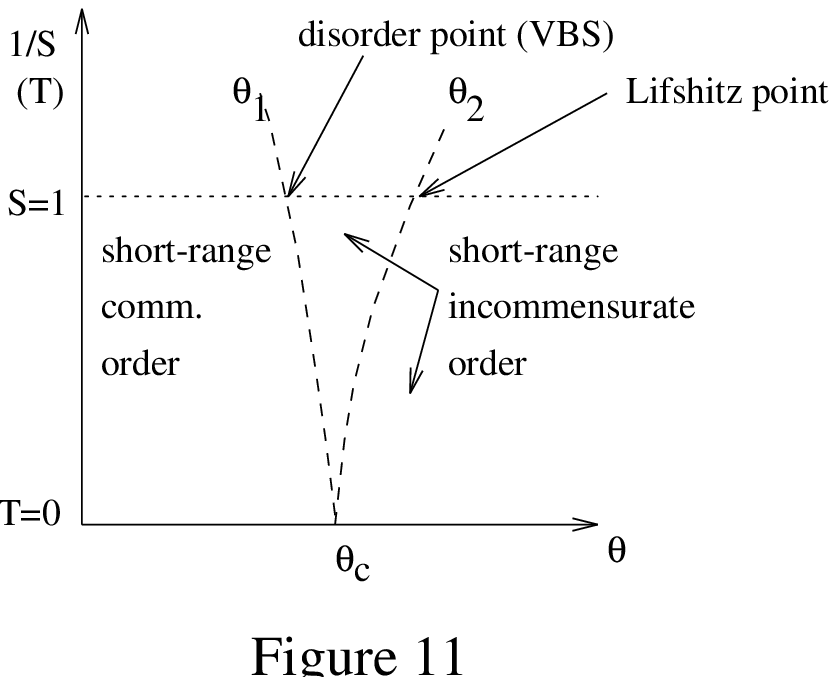}}
\vfill
\eject
\bye

\vfill
\eject
\bye